# *Topological origin of electromagnetic energy sinks*

David E. Fernandes[1*], Mário G. Silveirinha[1,2†]

[1]*University of Coimbra, Department of Electrical Engineering – Instituto de Telecomunicações, 3030-290 Coimbra, Portugal*

[2]*University of Lisbon, Instituto Superior Técnico, Avenida Rovisco Pais, 1, 1049-001 Lisboa, Portugal*

## Abstract

It was recently highlighted that nonreciprocal surface waves can be stopped in suitably designed waveguides, leading to the formation of hotspots wherein the electromagnetic fields are massively enhanced. Here, we prove that this phenomenon has a topological origin and is related to a breakdown of the bulk-edge correspondence in electromagnetic continua with no spatial cut-off. Our theoretical analysis shows that nonreciprocal electromagnetic continua can used to realize energy sinks with ultra-singular fields that absorb all the energy generated in their surroundings. Moreover, it is proven that similar energy sinks may be formed in fully reciprocal optical platforms with a parity-time-duality ($\mathcal{P}\cdot\mathcal{T}\cdot\mathcal{D}$) symmetry.

---

[*] E-mail: dfernandes@co.it.pt

[†] To whom correspondence should be addressed: E-mail: mario.silveirinha@co.it.pt

# I. Introduction

The use of topological ideas is one of the most exciting theoretical developments in optics in recent years [1, 2]. The dawn of topological photonics may be linked to the pioneering work of Haldane and Raghu [1] where it was shown that the different topological phases of a photonic crystal can be classified by integer numbers, known as Chern invariants. The topological invariants are determined by the photonic band structure of the material. A gap Chern number can be assigned to each band-gap [1]. Notably, topological materials may support back-scattering immune wave propagation at their boundaries, even though they do not support bulk states. This remarkable result has led to the development of many topological-inspired photonic systems [1-19].

The net number of unidirectional edge modes in a topological system is determined by the bulk edge correspondence principle [3, 20, 21, 22]. Importantly, the topological classification is usually restricted to materials with an intrinsic periodicity, for instance a photonic crystal [23], as it is necessary that the underlying wave vector space is a closed surface with no boundaries [24]. In electromagnetic continua the wave vector (**k**) domain is the unbounded Euclidean space, and hence generically such structures cannot be topologically classified. Notably, it was recently demonstrated that it is possible to characterize the topological phases of a wide class of bianisotropic continua [24] with a spatial cut-off $k_{\max}$ that guarantees that when $k \to \infty$ the material response becomes reciprocal (and asymptotically the same as that of a standard dielectric).

Yet, the response of most conventional materials is to an excellent approximation local, i.e., independent of the wave vector, and thereby typical material models do not include an explicit cutoff. In Ref. [21] it was shown that the role of the spatial cut-off can be imitated in practice by separating the relevant materials with a small air gap. Specifically, a tiny air gap with thickness $d$ may accurately mimic a cutoff $k_{\max} \sim 1/d$ [21]. Under this framework, the bulk-edge correspondence can be used to predict the number of unidirectional edge-states in conventional systems formed electromagnetic media with no intrinsic periodicity [21]. Therefore, the application of topological methods to electromagnetic continua is subtler than in photonic crystals, as it is essential to ensure that the spatial cut-off is implemented in the platform of interest, e.g., with a small air gap or alternatively taking into account charge diffusion effects [25, 26].

Previous works demonstrated the possibility of concentrating electromagnetic energy in tight regions of space using either tapered plasmonic waveguides [27-29] or nonreciprocal surface waves [30-39]. Other types of electromagnetic energy sinks based on super-absorbing photonic platforms have also been proposed recently [40-42]. Most remarkably, it has been shown that in some conditions it is possible to completely stop a propagating nonreciprocal surface wave and that this effect may lead to the formation of hotspots where the electromagnetic fields are greatly enhanced [35-39].

Here, we highlight for the first time that this phenomenon has a topological origin and is due to a breakdown of the bulk-edge correspondence [21]. Consistent with Refs. [35-39], we find that nonreciprocal materials without a spatial cut-off (or more generally with a weak spatial dispersion) enable the formation of energy sinks that take in all the energy generated in the surroundings. Different from conventional absorbers where the dissipation is distributed in space, in an energy sink the absorption occurs in an extremely tight region of space and is free of back-scattering. Our theoretical analysis shows that in a pumped scenario and in the hypothetical limit of no material absorption, the system cannot reach a stationary state and that the energy stored in the "sink" may grow steadily with time. In realistic systems, the energy accumulated in the sink is dissipated in the form of heat and the fields may be characterized by ultra-singularities, i.e., have such a strong divergent behavior that the stored energy density is "almost" non-integrable. Furthermore, relying on the concept of parity-time-duality ($\mathcal{P}\cdot\mathcal{T}\cdot\mathcal{D}$) invariance [18], it is demonstrated that similar electromagnetic sinks may be formed in time-reversal invariant (reciprocal) systems. Specifically, we propose for the first time a route to realize reciprocal energy sinks that can drag towards their interiors *all* the energy generated in the surrounding environment with no reflections, leading to full absorption in a very tight region of space. Electromagnetic sinks may be useful for energy harvesting [43, 44] or to enhance nonlinear effects, e.g., in higher harmonic generation and frequency conversion [45-48].

This article is organized as follows. In Section II, it is highlighted that the bulk-edge correspondence breaks down in electromagnetic continua. It is proven that backscattering immune edge states offer the opportunity to greatly enhance the strength of the electromagnetic fields. In Section III, it is shown that nonreciprocal effects may lead to the formation of ultra-singular electromagnetic sinks in systems with no wave vector cut-off. Furthermore, we propose a solution to have similar "sinks" in time-reversal invariant optical platforms. Finally, in Section IV the conclusions are drawn. The time dependence of the electromagnetic fields is assumed of the type $e^{-i\omega t}$.

## II. Breakdown of the bulk-edge correspondence

To illustrate the peculiarities of nonreciprocal systems formed by electromagnetic continua, we consider an edge-type waveguide formed by a gyrotropic material half-space and a metal half-space, separated by an air gap with thickness *d.* The structure is invariant along the *z*-direction, as shown in Fig. 1.

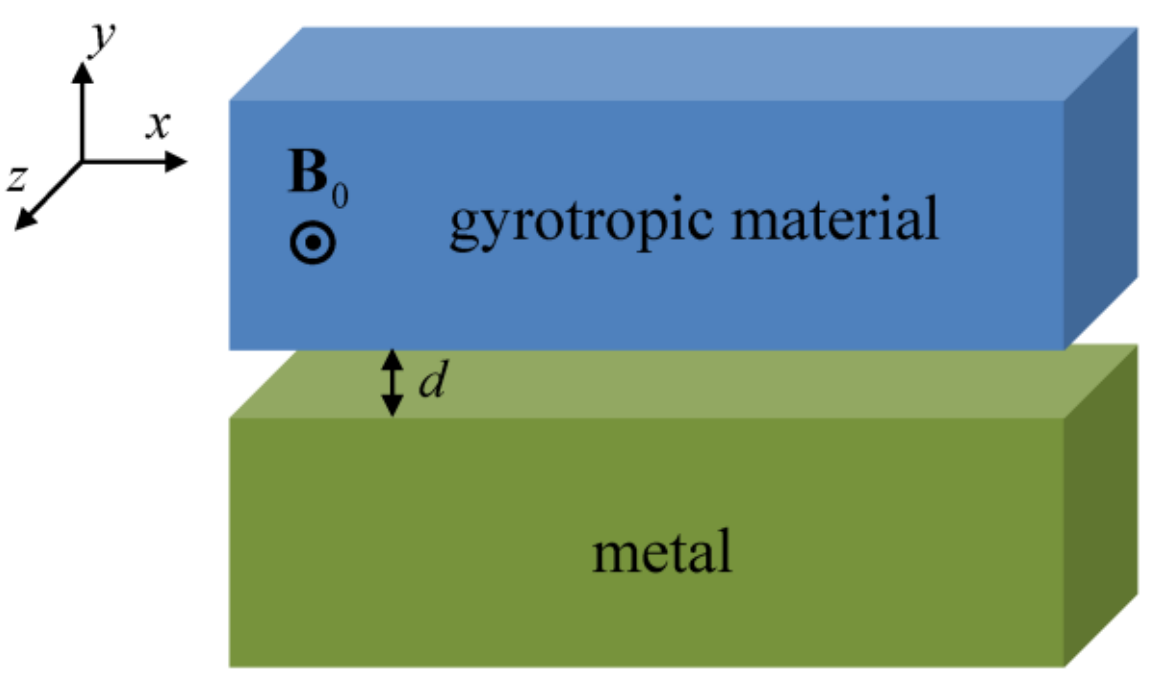

Fig. 1 Geometry of the interface between a gyrotropic material half-space and a metal half-space separated by an air gap with thickness *d*.

The gyrotropic material is modeled by the relative permittivity tensor $\bar{\bar{\varepsilon}} = \varepsilon_t \mathbf{1}_t + i\varepsilon_g \hat{\mathbf{z}} \times \mathbf{1}_t + \varepsilon_a \hat{\mathbf{z}} \otimes \hat{\mathbf{z}}$, where $\mathbf{1}$ is the identity matrix and $\mathbf{1}_t = \mathbf{1} - \hat{\mathbf{z}} \otimes \hat{\mathbf{z}}$. The nontrivial elements of the permittivity are $\varepsilon_{12} = -\varepsilon_{21} = -i\varepsilon_g$, $\varepsilon_{11} = \varepsilon_{22} = \varepsilon_t$ and $\varepsilon_{33} = \varepsilon_a$. Note that the $\hat{\mathbf{z}} \times \mathbf{1}_t$ represents the matrix $\begin{pmatrix} 0 & -1 & 0 \\ 1 & 0 & 0 \\ 0 & 0 & 0 \end{pmatrix}$. For a magnetized plasma (e.g., a magnetized semiconductor) the permittivity components satisfy [49]:

$$\varepsilon_t = 1 - \frac{\omega_p^2 (\omega + i\Gamma)}{\omega (\omega + i\Gamma)^2 - \omega\omega_0^2}, \quad \varepsilon_g = \frac{\omega_p^2 \omega_0}{\omega\omega_0^2 - \omega (\omega + i\Gamma)^2}, \quad \varepsilon_a = 1 - \frac{\omega_p^2}{\omega (\omega + i\Gamma)}. \quad (1)$$

with $\omega_p$ the plasma angular frequency, $\omega_0 = -qB_z / m = +eB_z / m$ the cyclotron frequency, $\mathbf{B}_0 = +B_z \hat{\mathbf{z}}$ is the static magnetic bias, and $\Gamma$ is the collision frequency associated with the material loss. The metal is described by a Drude dispersion model with the relative permittivity given by $\varepsilon_m(\omega) = 1 - \frac{\omega_{pm}^2}{\omega (\omega + i\Gamma_m)}$, where $\omega_{pm}$ and $\Gamma_m$ are the plasma and collision frequencies, respectively.

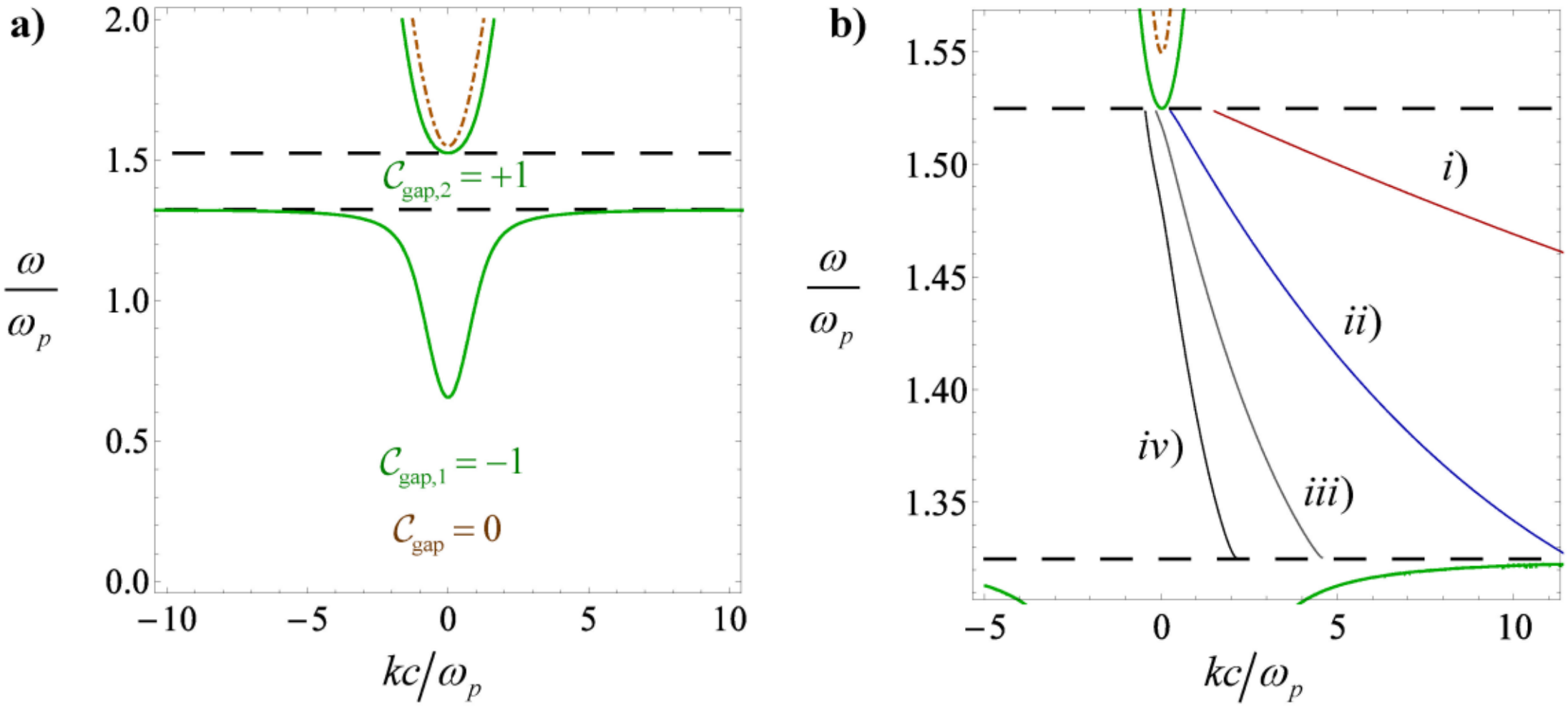

Fig. 2 **a)** Band diagram and gap Chern numbers of a lossless gyrotropic material with $\omega_0 = 0.87\omega_p$ (green solid lines) and of a lossless Drude plasma with $\omega_{pm} = 1.55\omega_p$ (brown dot-dashed line). The high-frequency band-gap common to both materials is highlighted with dashed black curves. **b)** Edge modes dispersion at a gyrotropic material-metal interface with an air gap with thickness: *(i)* $d = 0.025c/\omega_p$ (red curve); *(ii)* $d = 0.1c/\omega_p$ (blue curve); *(iii)* $d = 0.25c/\omega_p$ (gray curve); *(iv)* $d = 0.5c/\omega_p$ (black curve). The remaining curves are defined as in **a**).

Figure 2a shows the photonic band diagram of the transverse magnetic (TM) polarized eigenwaves (with $H_z \neq 0$, $E_z = 0$ and $\partial/\partial z = 0$) and the corresponding gap Chern numbers for a magnetized plasma with $\omega_0 = 0.87\omega_p$ and $\Gamma = 0^+$, and for a Drude plasma with $\omega_{pm} = 1.55\omega_p$ and $\Gamma_m = 0^+$ (except if explicitly stated otherwise, these structural parameters are fixed from hereon). For the gyrotropic material, the dispersion of the bulk modes (propagation in the *xoy* plane is implicit) is determined by $\varepsilon_{ef}\left(\omega/c\right)^2 = k_x^2 + k_y^2$, with $c$ is the light speed in a vacuum and $\varepsilon_{ef} = \left(\varepsilon_t^2 - \varepsilon_g^2\right)/\varepsilon_t$ [24]. The gap Chern numbers are found by enforcing a high-frequency cut-off in the material response, as explained in Refs. [24, 50]. The results of Fig. 2a confirm that the magnetized plasma has non-zero gap Chern numbers and hence is topologically nontrivial. In contrast, the gap Chern number of the metal is zero, and therefore the material is topologically trivial.

From the bulk-edge correspondence [3, 21, 22], one may expect that the planar interface of the two materials supports a single unidirectional edge state in each of the band-gaps ($\delta\mathcal{C}=\left|\mathcal{C}_{\text{gap,gyro}}-\mathcal{C}_{\text{gap,Drude}}\right|=1$). Next, we focus on the high frequency band-gap of the magnetized plasma for which $\mathcal{C}_{\text{gap,2}}=+1$. This band-gap is determined by $\omega_L<\omega<\omega_H$, with $\omega_H=\frac{|\omega_0|}{2}+\sqrt{\left(\frac{\omega_0}{2}\right)^2+\omega_p^2}$ and $\omega_L=\sqrt{\omega_0^2+\omega_p^2}$. Figure 2b shows the dispersion of the edge states for different values of *d* (see Appendix A for the calculation details). Consistent with the bulk-edge correspondence, for a finite (and arbitrarily small) gap *d* the system supports a single unidirectional edge state. Notably, the unidirectional mode is a backward wave with a negative group velocity $v_{\text{g},x}=\partial\omega_k/\partial k_x<0$ (the energy flows along the –*x*-direction) and a positive phase velocity $v_{\text{p}}=\omega_k/k_x>0$. Crucially, as the separation between the materials decreases ($d\rightarrow 0$) the edge mode dispersion becomes flatter and the group velocity of the edge state approaches zero (see the curve *i* in Fig. 2b calculated for $d=0.025\,c/\omega_p$). Furthermore, when the gap is removed, the material interface does not support edge modes in the common band-gap and thereby the bulk-edge correspondence breaks down when the spatial cut-off is not enforced (in the present example, the spatial cut-off is determined by the air-gap between the materials $k_{\max}\sim 1/d$) [21] (see Ref. [51] for another example where the group velocity of an edge mode vanishes when $d\rightarrow 0$).

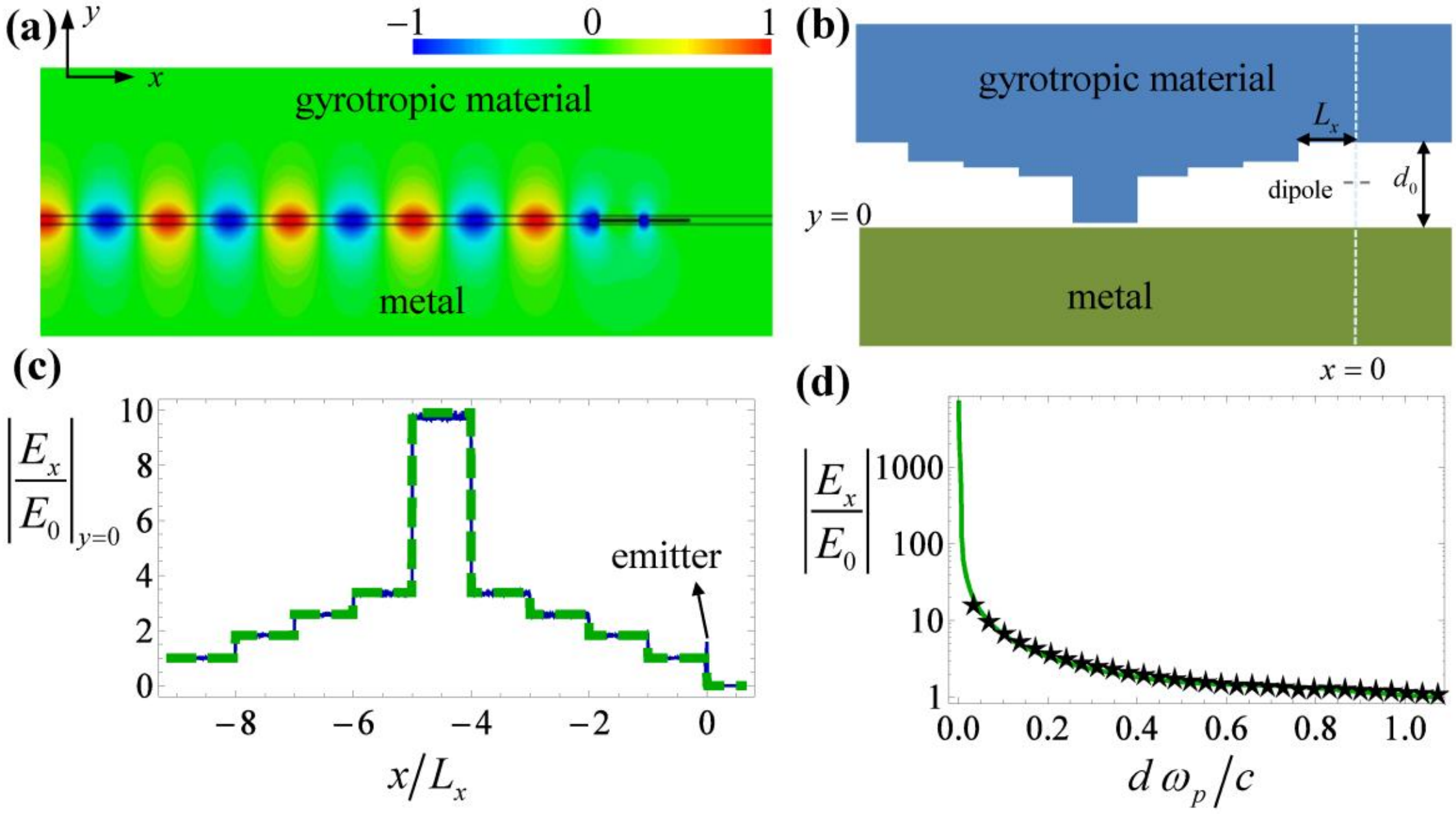

Fig. 3. **(a)** Time snapshot of the electric field $E_x$ radiated by a horizontal electric dipole placed at the middle of an air gap separating the gyrotropic material and the metal. The thickness of the air gap is $d = 0.25\,c/\omega_p$ and the oscillation frequency is $\omega = 1.45\omega_p$. **(b)** Side view (cut in the *xoy* plane) of a tapered topological waveguide uniform along *z*. The gap thickness is tapered from $d = d_0 \equiv 1.07\,c/\omega_p$ down to $d = 0.07\,c/\omega_p$. **(c)** Normalized amplitude of the electric field $\left|E_x/E_0\right|$ at the interface $y = 0$ between the air region and the metal in the tapered waveguide. Full wave simulation results: solid blue line; Theoretical results: dashed green line. **(d)** Amplitude of the electric field $\left|E_x/E_0\right|$ as a function of the air gap thickness. Full wave simulation results: black stars; Theoretical results: solid green line.

Figure 3a shows a time snapshot of the electric field radiated by a short horizontal dipole placed at the middle of the air gap, calculated with CST-MWS [52]. The gap thickness is $d = 0.25\,c/\omega_p$ and the oscillation frequency of the dipole ($\omega = 1.45\omega_p$) lies in the relevant band-gap. Consistent with the bulk-edge correspondence and Fig. 2b*iii*, the dipole excites a single edge mode in the waveguide. The edge mode propagates along the –*x*-direction but with a positive phase velocity (the time animation of the emitted field can be found in the Supplemental Material [53]), so that it corresponds to a backward wave.

Next, we analyze how the tapering of the air gap (see Fig. 3b) can affect the unidirectional edge mode propagation in the waveguide. From Fig. 3a, the fields are concentrated in the air gap; hence, one may expect a strong field enhancement when $d \to 0$ because the edge mode is immune to backscattering, i.e., it cannot be reflected back to the source and it cannot leak to the bulk regions. The results depicted in Fig.3c confirm this idea by showing that the electric field $|E_x|$ may be significantly enhanced when the separation between materials is reduced. The electric field enhancement is roughly 10 times when the gap thickness is reduced by the same factor. The amplitude of the electric field is nearly constant in each uniform section of the waveguide (not shown).

The field enhancement can be precisely predicted using analytical methods. Indeed, due to the absence of back-scattering, the power flow along the *x*-direction must be independent of *x* in the absence of material loss, i.e., $P = \int S_x dy = const.$ (the power flow is normalized to the length along *z*). The fields of an edge mode ($H_z(x, y; d)$, $E_x(x, y; d)$ and $E_y(x, y; d)$) can be analytically found for a fixed *d* (see Appendix A). Using the adiabatic approximation, it follows that for a tapered waveguide $H_z \sim H_z(x, y; d(x))$, etc, with $d(x)$ the function that determines the air gap profile. For each *d* (i.e., in each uniform section of the waveguide), the overall scaling factor of the edge mode amplitude is determined by imposing $P_0 = \frac{1}{2}\int_{-\infty}^{+\infty} \mathrm{Re}\{E_y(y; d) H_z^*(y; d)\} dy$, with $P_0$ the power emitted by the dipole. The result obtained with the adiabatic approximation is represented in Fig. 3c (dashed-green line). Notwithstanding the complexity of the tapered waveguide, the theory agrees almost exactly with the full wave simulation. The reason is that the inexistence of scattering channels makes the adiabatic approximation almost exact. Indeed, because of the conservation of energy, the amplitude of the fields in a uniform section of the waveguide (with a constant $d$ and sufficiently far from the abrupt transitions) must be independent of the considered tapering.

The electric field $|E_x|$ at the metal-air interface was calculated as a function of the air gap thickness (Fig. 3d). The thickness range for the full wave results (discrete symbols) is restricted to the minimum value $d = c/(30\omega_p)$ due to computational limitations. In agreement with the previous discussion, as $d \to 0$ the electric field is greatly enhanced and is roughly inversely proportional to $1/d$. The magnetic field amplitude in the gap is nearly insensitive to $d$ (not shown), so that the growth rate of the electric field is consistent with the conservation of energy ($P = \int S_x dy = const.$).

## III. Topological electromagnetic sinks

### *A. Nonreciprocal energy sink*

The analysis of the previous section raises an obvious question: "what happens if the cross section of the waveguide is modified so that the gap thickness is abruptly reduced down to precisely zero"? (see Fig. 4a). Remarkably, such a scenario creates a deadlock for the guided mode propagating in the "input" waveguide (guide with the finite gap thickness $d_0 = 0.5c/\omega_p$ in Fig. 4a). At the junction point, the incoming wave has no place to go: it cannot propagate in the "output" waveguide (with $d = 0$), it cannot escape to the bulk regions and it cannot be back-reflected because the edge mode is immune to backscattering. Therefore, in the ideal case of lossless materials, the electromagnetic energy is forced to accumulate at the connection point between the waveguides, which behaves as an energy sink. This phenomenon was previously discussed in the literature [35-39], but its topological origin, unveiled here by our analysis, was unnoticed. Furthermore, it is known since the 50's of last century that similar "propagation deadlocks" can be created in closed metallic waveguides partially filled with ferrites [30-33]. In particular, these pioneering studies highlighted that a wave cannot be stopped without material absorption because this would violate the third law of thermodynamics [30-33].

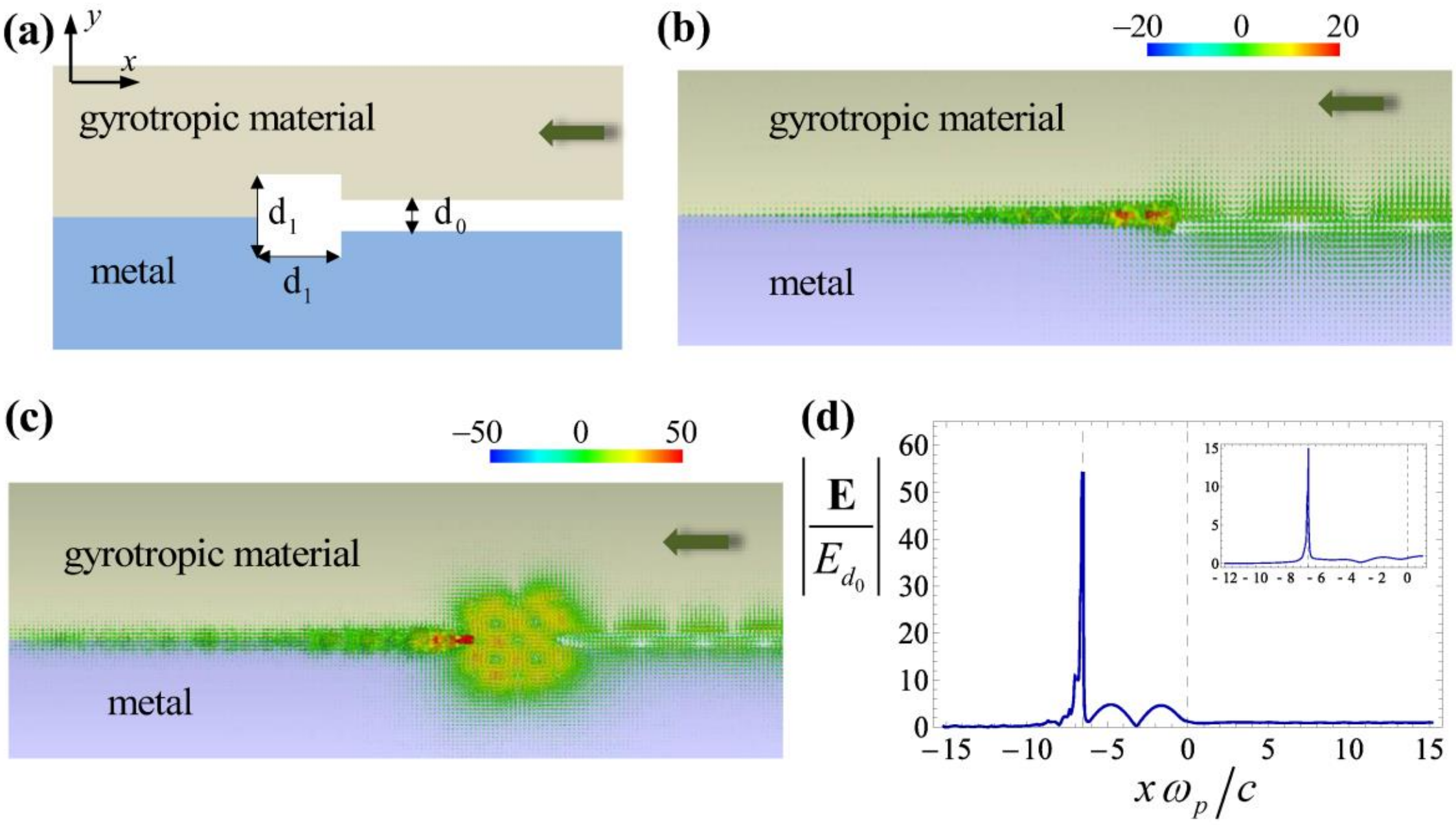

Fig. 4. **(a)** Junction between two edge-type waveguides. The input waveguide (at the right-hand side) is formed by a gyrotropic material and a metal separated by an air gap with thickness $d_0 = 0.5\, c/\omega_p$. The output waveguide (at the left-hand side) is formed by the same materials but the air gap is suppressed. A square cavity with dimensions $d_1 \times d_1$ connects the two regions. The collision frequencies of the gyrotropic material and metal are $\Gamma/\omega_p = 10^{-4}$ and $\Gamma_m/\omega_{pm} = 3\times10^{-3}$, respectively. **(b)** Time snapshot of the electric field for a direct connection of the guides ( $d_1 = 0$ ) and $\omega = 1.45\omega_p$. **(c)** similar to **(b)** but the two guides are connected by a square cavity with $d_1\omega_p/c \approx 6.53$. **(d)** Amplitude of the electric field along the line *y*=const that includes the gyrotropic material-metal interface in the setup **(c)**. The cavity limits are marked with dashed vertical lines. The inset shows the effect of realistic loss ( $\Gamma/\omega_p = 0.11$, similar to InSb) in the field enhancement.

First, we numerically illustrate the emergence of an energy "sink" in systems formed by dissipative materials using full wave simulations. Figure 4 shows time snapshots of the stationary electric field for a time harmonic excitation when the input and output guides are directly connected, $d_1 = 0$ (Fig. 4b), and when the two guides are coupled by a square cavity (Fig. 4c). The input guide (on the right-hand side) is fed by an edge mode that propagates along the $-x$-direction. The materials are the same as in Fig. 2. As seen, in both examples an electric field sink is created at the junction of the waveguides. The electric field intensity is depicted in Fig. 4d for the example of Fig. 4c and reveals an amplitude enhancement on the order of 50. The enhancement factor of the electric field in the sink depends on the loss level (see the inset of Fig. 4d for a loss strength comparable to that of InSb [54, 55]). The field enhancement can be arbitrarily large when the dissipation terms approach zero (see Fig. 3d; here, we note that when the material loss is small it is challenging to obtain fully converged numerical results because of the ultra-singular behavior of the fields near the junction). Due to the absence of scattering channels, all the incoming energy is dragged into the sink where it is dissipated. Evidently, when the system has additional lossy elements, e.g., if a lossy nanoparticle is placed near the sink, part of the energy may be dissipated outside the sink. Yet, even in this case there are no back-reflections.

Applying the Poynting theorem to some volumetric region $V$ enclosing the junction we get $\frac{d\mathcal{E}}{dt} + p_d = p_S$, with $\mathcal{E}$ the energy stored in $V$, $p_S$ the net power flowing inside $V$ through the surface boundary, and $p_d$ the power dissipated in $V$ due to material loss. In general, $p_S = p_{\text{in}} - p_{\text{out}}$ with $p_{\text{in}}$ the power transported by the incident edge wave and $p_{\text{out}}$ the power coupled to the available scattering channels. In the absence of scattering channels (as in Fig. 4), $p_S = p_{\text{in}} > 0$ is strictly positive. Let us suppose that the incident wave has a spectrum peaked at some frequency $\omega$. The dissipated power may be assumed of the form $p_d = \Gamma_{\text{diss}}\mathcal{E}$ with $\Gamma_{\text{diss}}$ the dissipation rate at the considered frequency due to material absorption. Hence, in our system the time evolution of the stored energy is determined by:

$$\frac{d\mathcal{E}}{dt}+\Gamma_{\text{diss}}\mathcal{E} = p_{\text{in}} \qquad (2)$$

With material loss, it is possible to reach a stationary-state when the excitation is time-harmonic. The stored energy in the steady-state ($d/dt=0$) is $\mathcal{E} = p_{\text{in}}/\Gamma_{\text{diss}}$ and notably *diverges* in the limit $\Gamma_{\text{diss}} \to 0^+$ (see also [33]). Note that this situation is quite peculiar and is *not* observed in conventional electromagnetic systems. Indeed, in standard systems as $\Gamma_{\text{diss}} \to 0^+$ the incoming wave is invariably scattered away from the junction such that $p_{\text{out}} - p_{\text{in}} \to 0^+$. Thereby, in conventional platforms the stationary-state energy $\mathcal{E} = \left(p_{\text{in}} - p_{\text{out}}\right)/\Gamma_{\text{diss}}$ remains finite as the material loss is suppressed ($\Gamma_{\text{diss}} \to 0^+$). Thus, the system of Fig. 4 is quite unique as in the limit of no material loss it can accumulate all the incoming energy at the vertex.

Furthermore, when $\Gamma_{\text{diss}}=0$ we see from Eq. (2) that $\mathcal{E} = \mathcal{E}_0 + p_{\text{in}}t$, i.e., the energy grows linearly with time under a time-harmonic excitation, similar to a lossless LC circuit excited at the resonant frequency. The electromagnetic sink has a topological origin, as it is due to the breakdown of the bulk-edge correspondence when the wave vector cut-off is suppressed.

It should be noted that realistic materials may have some high-frequency spatial cut-off due to their granular nature [21] or due to charge diffusion effects [25]. In principle, such a cut-off ensures that the waveguide in Fig. 4a supports edge states even when $d=0$ [21, 25]. In that case, the characteristic size of the edge states is on the order of the atomic lattice constant [21] or of some characteristic diffusion length [25]. Thus, in realistic situations the edge states are extremely confined at the interface when $d=0$ [21], and thus will be strongly damped by unavoidable material loss. Therefore, a hypothetical high-frequency spatial cut-off due to the material granularity does not preclude the formation of the energy sinks discussed here.

### *B. "Wedge" energy sink*

To further illustrate the peculiar properties of the topological energy sink, next we present a quasi-static analysis of the field singularities at the vertex of a nonreciprocal "wedge" with the geometry of Fig. 5 (a related study was done by Ishimaru in [33], but without the quasi-static approximation). The angular sector $0<\varphi<\pi/2$ is filled with an electric gyrotropic material and the surrounding walls are *i)* a perfect electric conductor (PEC) at $\varphi=0$ and *ii)* a perfect magnetic conductor (PMC) at $\varphi=\pi/2$. It can be shown that the PMC-gyrotropic material interface does not support any edge state, while the PEC-gyrotropic material interface supports a single edge state (even without an air gap in between the materials).

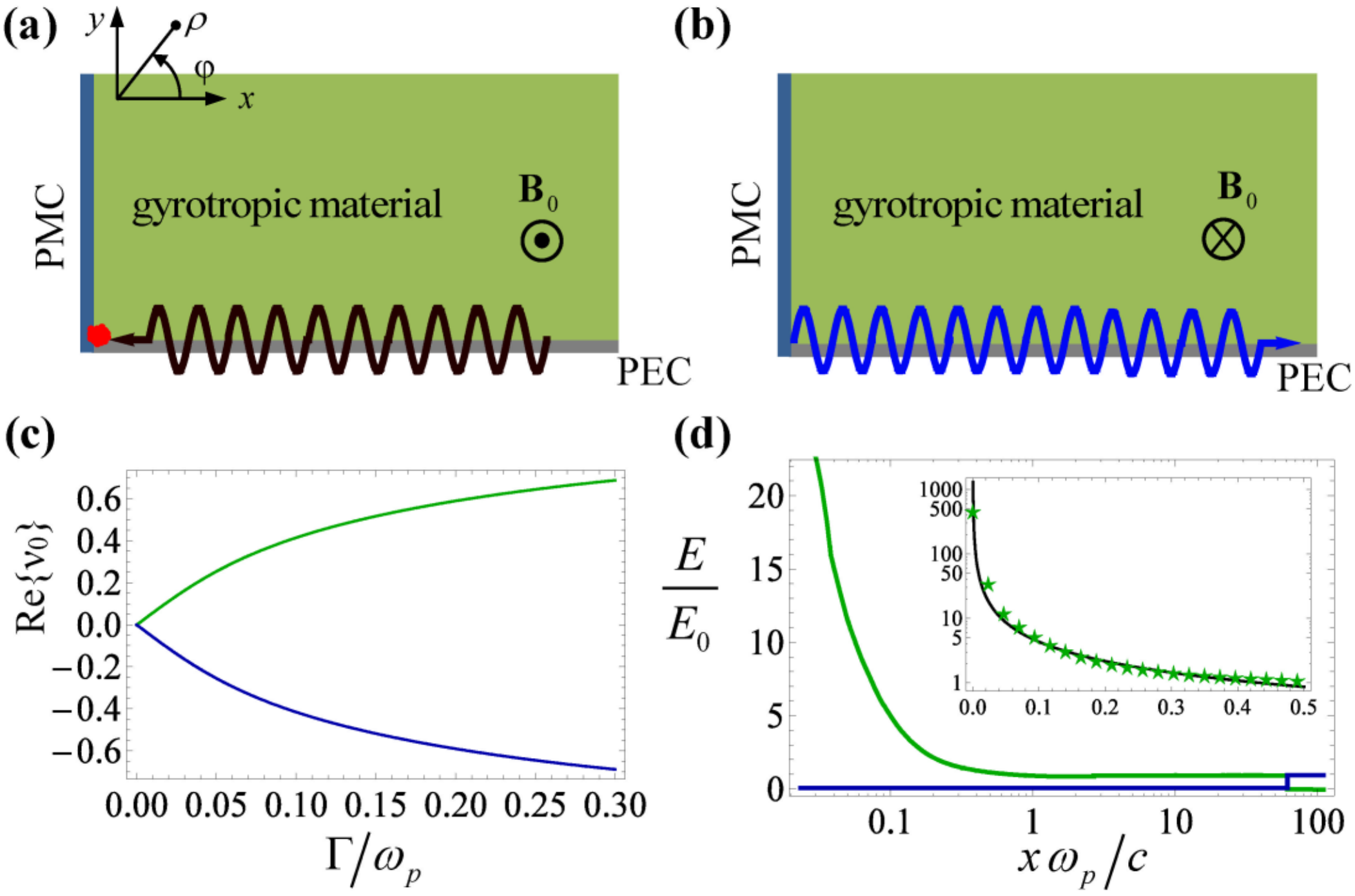

Fig. 5. Gyrotropic material wedge delimited by PEC and PMC walls. The edge mode at the PEC boundary flows **(a)** towards from the wedge vertex and **(b)** away the wedge vertex, depending on the magnetic bias orientation. **(c)** Real part of $\nu_0$ as a function of the collision frequency $\Gamma$ of the magnetized plasma. Green curve: $\mathbf{B}_0 = +B_z\hat{\mathbf{z}}$ ($\omega_0 > 0$); Blue curve: $\mathbf{B}_0 = -B_z\hat{\mathbf{z}}$ ($\omega_0 < 0$). **(d)** Normalized electric field amplitude $E/E_0$ at the PEC interface when the structure is excited with a small emitter. The collision frequency is $\Gamma/\omega_p = 1\times10^{-4}$. Green curve: $\mathbf{B}_0 = +B_z\hat{\mathbf{z}}$; Blue curve: $\mathbf{B}_0 = -B_z\hat{\mathbf{z}}$. The inset shows a close-up of the field intensity near the wedge vertex; green stars: full wave simulator results; black line: quasi-static model results.

Under a quasi-static approximation, the time-harmonic electromagnetic fields may be assumed of the form $\mathbf{E} = -\nabla\phi$ and $\mathbf{H} \approx 0$ with $\phi$ the electric potential. Since the electric displacement vector has zero divergence, $\nabla\cdot\mathbf{D} = 0$, the electric potential satisfies a Laplace-type equation $\nabla\cdot\left(\overline{\overline{\varepsilon}}\cdot\nabla\phi\right) = 0$. We look for solutions of the form $\phi = \rho^{\nu} g(\varphi)$, with $\rho$ the radial distance to the wedge vertex, $g(\varphi)$ some function that describes the azimuthal variation of the potential and $\nu$ an unknown constant. Using $\mathbf{D} = -\overline{\overline{\varepsilon}}\cdot\nabla\phi$, it is found that $\mathbf{D} = -\rho^{\nu-1}\left(\varepsilon_{11}\nu g + \varepsilon_{12}\partial_{\varphi} g\right)\hat{\boldsymbol{\rho}} - \rho^{\nu-1}\left(\varepsilon_{11}\partial_{\varphi} g - \varepsilon_{12}\nu g\right)\hat{\boldsymbol{\varphi}}$ with $\partial_{\varphi} = \partial/\partial\varphi$. Hence, the electric potential must satisfy $\nu\left(\varepsilon_{11}\nu g + \varepsilon_{12}\partial_{\varphi} g\right) + \partial_{\varphi}\left(\varepsilon_{11}\partial_{\varphi} g - \varepsilon_{12}\nu g\right) = 0$, which in the gyrotropic sector (wherein the permittivity tensor elements are constant) reduces to $\partial_{\varphi}^2 g + \nu^2 g = 0$. This proves that $g$ is a linear combination of the functions $\cos(\nu\varphi)$ and $\sin(\nu\varphi)$.

The boundary conditions require that $\phi = 0$ at the PEC wall ($\varphi = 0$) and $\mathbf{D}\cdot\hat{\boldsymbol{\varphi}} = 0$ at the PMC wall ($\varphi = \pi/2$). From the former boundary condition, it follows that $g(\varphi) = C\sin(\nu\varphi)$ ($C$ is some arbitrary constant); hence the electric potential in the gyrotropic sector is of the form:

$$\phi = C\rho^{\nu}\sin(\nu\varphi), \quad 0<\varphi<\frac{\pi}{2}. \tag{3}$$

The boundary condition at the PMC interface leads to $\cot\left(\nu\frac{\pi}{2}\right)=-i\varepsilon_g/\varepsilon_t$, where we used $\varepsilon_{12}=-i\varepsilon_g$ and $\varepsilon_{11}=\varepsilon_t$. There are multiple solutions for $\nu$ determined by:

$$\nu_n=\frac{2}{\pi}\left(i\,\mathrm{arccoth}\frac{\varepsilon_g}{\varepsilon_t}+n\pi\right),\quad n=0,\pm1,... \tag{4}$$

Importantly, not all values of *n* correspond to physical solutions. Indeed, in order that in a stationary-state the electromagnetic energy stored near the vertex remains finite, it is necessary that $|\mathbf{E}|^2\rho\sim\rho^{2\mathrm{Re}\{\nu\}-2+1}=\rho^{2\mathrm{Re}\{\nu\}-1}$ be square-integrable (near the vertex). This is only possible if $\mathrm{Re}\{\nu\}>0$.

In the relevant band-gap of the gyrotropic material ($\omega_L<\omega<\omega_H$) it may be easily checked that $\varepsilon_t>0$ and thereby $|\varepsilon_g|>\varepsilon_t$ (because $\varepsilon_{ef}<0$ must be negative in a band-gap). Hence, in the limit of vanishing loss the term $\mathrm{arccoth}\frac{\varepsilon_g}{\varepsilon_t}$ in Eq. (4) is real-valued because $|\varepsilon_g/\varepsilon_t|>1$. This proves that without material loss the solution with *n*=0 ($\nu_0$ ) is a pure imaginary number ($\mathrm{Re}\{\nu\}=0$), and hence is forbidden in a stationary regime.

Any amount of material loss endows $\nu_0$ with a real part, which, if positive, allows the energy density to be integrable near the vertex and the *n*=0 solution to be physically admissible. The sign of $\mathrm{Re}\{\nu_0\}$ depends on $\varepsilon_g$ and thereby on the orientation of the magnetic bias. Figure 5c depicts $\mathrm{Re}\{\nu_0\}$ as a function of the collision frequency $\Gamma$ of the magnetized plasma for the two possible orientations of the static magnetic bias. As seen, the real part of $\nu_0$ is positive only when the magnetic bias is oriented along the +*z*-direction ($\omega_0>0$). As illustrated in Fig. 5a, in this scenario the edge-state propagating at the PEC boundary flows towards the wedge vertex which behaves as an "energy sink". In contrast, for the opposite orientation of the magnetic bias (Fig. 5b), the edge mode flows away from the vertex, and there is no sink. In summary, as $\Gamma \to 0^+$ the fields near the vertex become ultra-singular only when $\omega_0 > 0$. In this case, the energy density of the *n*=0 solution is non-integrable in the $\Gamma = 0$ limit consistent with the fact that the energy sink does not support stationary solutions (the energy near the vertex grows linearly with time). The same singularity strength was analytically derived in Ref. [37], but for a ferrite-loaded waveguide.

Using a full wave simulator, we obtained the field radiated by a short dipole placed near the gyrotropic material-PEC interface for both orientations of the magnetic bias. The computed electric field profile is represented in Fig. 5d and confirms that the ultra-singularity occurs at the vertex only when $\mathbf{B}_0 = +B_z\hat{\mathbf{z}}$. When the magnetic field is biased in the opposite direction, the dipole excites an edge mode that propagates away from the wedge vertex and thereby does not create an electromagnetic field hotspot.

The quasi-static solution (3) yields an electric field of the form $\mathbf{E}\big|_{y=0} = E_y\hat{\mathbf{y}}\big|_{y=0} \sim \rho^{\nu-1}\hat{\mathbf{y}}$ at the PEC interface. The inset in Fig. 5d shows the superposition of the field profiles calculated using CST-MS [52] and with the quasi-static model (the overall scaling constant of the quasi-static model is adjusted by numerical fitting). As seen, the overall agreement is excellent, and the electric field enhancement at the "energy sink" is massive. The instantaneous electric field is given by $E_y(t)\big|_{y=0} = \mathrm{Re}\left\{E_y e^{-i\omega t}\right\} \sim \rho^{\mathrm{Re}\{\nu\}-1}\cos\left(\mathrm{Im}\{\nu\}\ln\rho - \omega t\right)$. The phase of the instantaneous field depends strongly on $\rho$, and thus the instantaneous field sign may be very sensitive to the distance to the vertex. However, since $\mathrm{Im}\{\nu\}$ is rather small this effect occurs only at very small distances.

To have an idea of the singularity strength in a realistic material, we calculated $\mathrm{Re}\{\nu\}$ for a magnetized InSb wedge. described by the gyrotropic permittivity model of Refs. [54, 55] (sample A of Ref. [54] with $\omega_p/2\pi = 5\,\mathrm{THz}$ and $\Gamma/\omega_p \approx 0.108$). This dispersive model is different from that in Eq. (1) because of the optical phonon contribution to the permittivity function [54]. Figure 6a represents $\left|\mathrm{Re}\{\nu\}\right|$ as a function of the frequency and of the bias magnetic field amplitude $B_0$. Since the electric field varies as $\rho^{\mathrm{Re}\{\nu\}-1}$ near the vertex, the strongest singularity occurs for small (positive) values of $\mathrm{Re}\{\nu\}$. Note that the sign of $\nu$ flips with the sign of the magnetic field. As seen in Fig. 6a, values of $\left|\mathrm{Re}\{\nu\}\right|$ near zero can be obtained with large values of $B_0$ and for $\omega/\omega_p << 1$. The detailed variation of $\left|\mathrm{Re}\{\nu\}\right|$ as a function of $B_0$ is shown in Fig. 6b for $\omega/\omega_p = 0.1$. For a moderate bias field $B_0 = 0.5\mathrm{T}$ the singularity strength is determined by $\mathrm{Re}\{\nu\} \approx 0.45$, and for a bias field as intense as $B_0 = 4\mathrm{T}$ by $\left|\mathrm{Re}\{\nu\}\right| \approx 0.07$. Hence, as the magnetic bias increases the singularity becomes more pronounced. We also did a similar study for a *n*-doped magnetized Si [55, 56] (not shown); due to the stronger material loss it is impractical to have $\left|\mathrm{Re}\{\nu\}\right|$ significantly less than one.

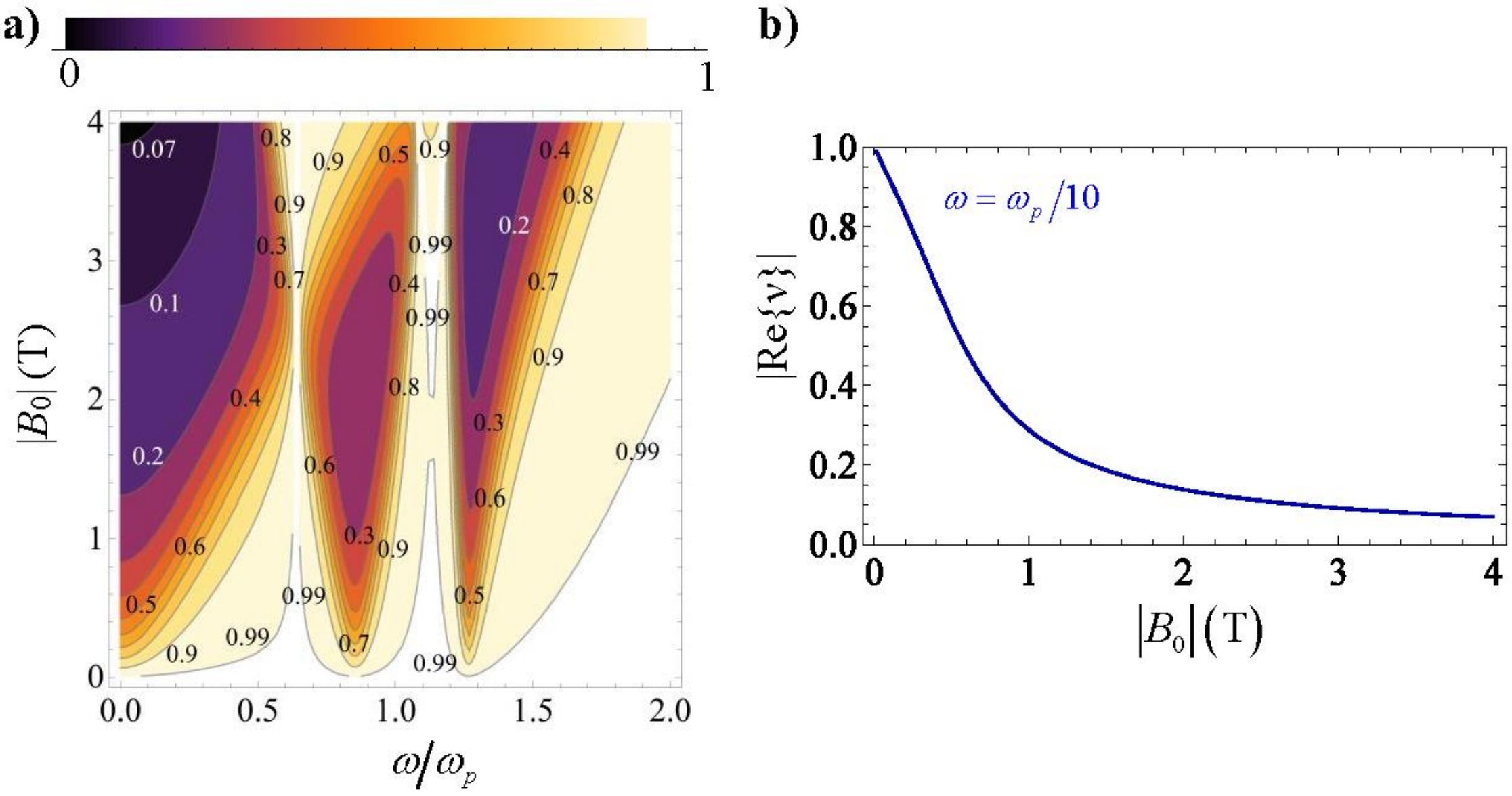

Fig. 6 **a)** Density plot of $\left|\mathrm{Re}\{v\}\right|$ as a function of the frequency and of the magnetic bias for InSb with $\omega_p/2\pi = 5$ THz and $\Gamma/\omega_p \approx 0.108$. **b)** Similar to **a)** but calculated at the fixed frequency $\omega = \omega_p/10$.

### *C. Reciprocal energy sink*

The results of the previous sections may suggest that the formation of "energy sinks" is only possible when the time-reversal symmetry is broken so that the propagation can be unidirectional. Surprisingly, we demonstrate in what follows that energy sinks can also be formed in *reciprocal systems*, which are inherently bi-directional.

As a starting point, we consider a modified wedge-structure (see Fig. 7a) such that the gyrotropic material has both an electric and a magnetic response. Specifically, it is supposed that the electric response is characterized by a relative permittivity tensor $\overline{\overline{\varepsilon}} = \varepsilon_t \mathbf{1}_t + i\varepsilon_g \hat{\mathbf{z}} \times \mathbf{1}_t + \varepsilon_a \hat{\mathbf{z}} \otimes \hat{\mathbf{z}}$ and that at the *frequency of interest* the magnetic response is determined by the relative permeability $\overline{\overline{\mu}} = \overline{\overline{\varepsilon}}^T = \varepsilon_t \mathbf{1}_t - i\varepsilon_g \hat{\mathbf{z}} \times \mathbf{1}_t + \varepsilon_a \hat{\mathbf{z}} \otimes \hat{\mathbf{z}}$. For simplicity, we assume that $\varepsilon_t$ and $\varepsilon_g$ follow the dispersive model of Eq. (1) with $\omega_0 > 0$ in the spectral range of interest, and that $\varepsilon_a = 1$.

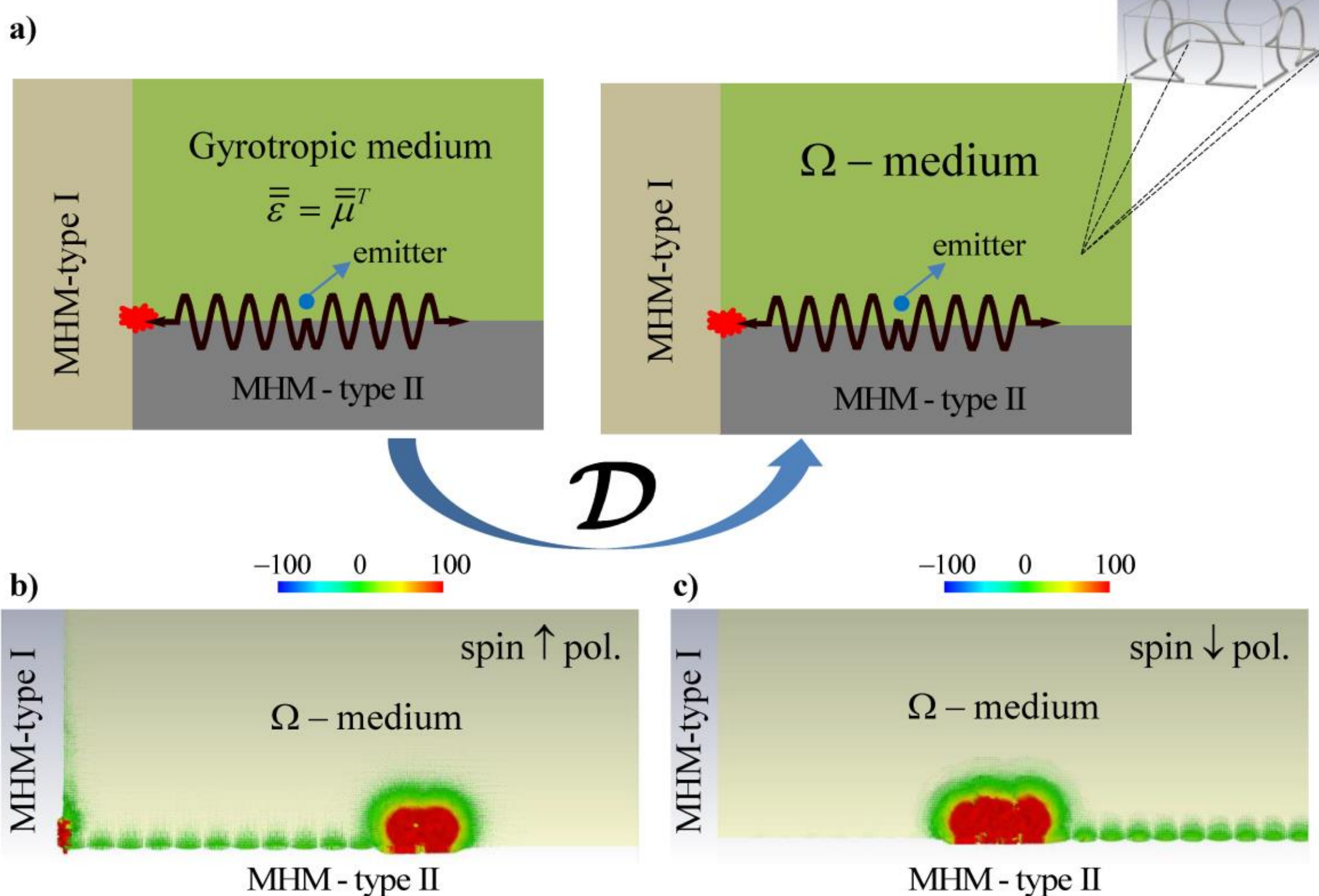

Fig. 7. **a)** A gyrotropic material with matched electrical and magnetic responses is surrounded by two matched hyperbolic metamaterials: a MHM-type I at the left-hand side lateral wall and a MHM-type II at the bottom wall. A duality mapping transforms the nonreciprocal platform into a *reciprocal* system with the same MHMs walls and the gyrotropic medium transformed into an $\Omega$-medium. **b)** Time snapshot of the spin up ($\uparrow$) polarized edge-mode, for $\omega = 1.45\omega_p$ excited by a small emitter of spin $\uparrow$ waves. **c)** Similar to **b)** but for a spin down ($\downarrow$) polarized edge-mode.

Furthermore, the PEC and PMC walls are replaced by *matched* hyperbolic metamaterials (MHMs), so that $\text{PMC} \rightarrow \text{MHM-I}$ and $\text{PEC} \rightarrow \text{MHM-II}$, with generic relative permittivity and permeability tensors $\bar{\bar{\varepsilon}}_i = \varepsilon_{\|,i}\left(\hat{\mathbf{x}}\hat{\mathbf{x}} + \hat{\mathbf{y}}\hat{\mathbf{y}}\right) + \varepsilon_{zz,i}\hat{\mathbf{z}}\hat{\mathbf{z}}$ and $\bar{\bar{\mu}}_i = \mu_{\|,i}\left(\hat{\mathbf{x}}\hat{\mathbf{x}} + \hat{\mathbf{y}}\hat{\mathbf{y}}\right) + \mu_{zz,i}\hat{\mathbf{z}}\hat{\mathbf{z}}$ ($i$=I,II) with $\bar{\bar{\varepsilon}}_i = \bar{\bar{\mu}}_i$ and

$$\varepsilon_{\|,\text{I}} = 1, \qquad \varepsilon_{zz,\text{I}} = -\infty, \qquad \text{(MHM-I)} \tag{5a}$$

$$\varepsilon_{\|,\text{II}} = -\infty, \qquad \varepsilon_{zz,\text{II}} = 1, \qquad \text{(MHM-II)} \tag{5b}$$

at the frequency of interest. Hyperbolic metamaterials are of great interest as they enable a plethora of exotic electromagnetic phenomena [57, 58]. Even though challenging, materials with the required properties (in the spectral range of interest) can in principle be synthesized using the metamaterial concept.

There are two relevant wave polarizations for propagation in the *xoy* plane: TM-polarization (with $H_z \neq 0$ and $E_z = 0$) and transverse electric (TE) polarization (with $H_z = 0$ and $E_z \neq 0$). Interestingly, the modified wedge structure has precisely the same electromagnetic response as the original wedge structure for TM-polarization (Fig. 5b). Indeed, for TM-polarization the relevant components of the hyperbolic metamaterials are $\varepsilon_{||,\mathrm{I}} = 1$, $\mu_{zz,\mathrm{I}} = -\infty$ (MHM-I behaves as a PMC) and $\varepsilon_{||,\mathrm{II}} = -\infty$, $\mu_{zz,\mathrm{II}} = 1$ (MHM-II behaves as a PEC). In addition, for TM-polarization the magnetic response of the gyrotropic material is trivial ($\mu_{zz} = \varepsilon_a = 1$). Therefore, the vertex of the modified wedge-structure absorbs all the energy that is transported by the TM edge mode at the MHM-II interface.

Importantly, the response to TE-polarized waves is dual of that for TM-waves. Indeed, for TE waves the relevant components of the hyperbolic metamaterials are $\mu_{||,\mathrm{I}} = 1$, $\varepsilon_{zz,\mathrm{I}} = -\infty$ (MHM-I behaves as a PEC) and $\mu_{||,\mathrm{II}} = -\infty$, $\varepsilon_{zz,\mathrm{II}} = 1$ (MHM-II behaves as a PMC). Furthermore, for TE-polarization the gyrotropic material has a trivial electric response $\varepsilon_{zz} = \varepsilon_a = 1$. It can be checked that the MHM-II interface now supports a TE-edge mode, but (because of the different signs of $\mu_{12}$ and $\varepsilon_{12}$) it propagates *away* from the vertex. The interface MHM-I does not support any TE edge modes, similar to TM-polarization.

In summary, the modified wedge structure *only* supports edge modes at the MHM-II (horizontal) interface. The TM-polarized edge mode propagates towards the vertex which behaves as a energy sink; in contrast, the TE-polarized edge mode flows away from the vertex and does not create energy singularities. Interestingly, as detailed in Ref. [18], the modified wedge structure is formed by parity-time-duality ($\mathcal{P}\cdot\mathcal{T}\cdot\mathcal{D}$) invariant materials. Waveguides invariant under the composition of the $\mathcal{P}\cdot\mathcal{T}\cdot\mathcal{D}$ operators are intrinsically bi-directional, consistent with the fact that the MHM-II – gyrotropic material interface supports edge modes that propagate along +*x* (TE-wave) and –*x* (TM-wave).

Remarkably, it was demonstrated in Ref. [18] that $\mathcal{P}\cdot\mathcal{T}\cdot\mathcal{D}$ invariant structures formed by *nonreciprocal* materials analogous to the ones considered here, can be mapped with a duality transformation into $\mathcal{P}\cdot\mathcal{T}\cdot\mathcal{D}$-invariant *reciprocal* structures. A generic duality transformation only acts on the electromagnetic fields, and does not affect either the space or time coordinates [59-61]. Due to this reason the wave phenomena in two optical platforms linked by a duality transformation are fundamentally equivalent. In our case, the duality transformation of interest is of the form $(\mathbf{E},\mathbf{H})\to(\mathbf{E}',\mathbf{H}')$ with [18]

$$\begin{pmatrix}\mathbf{E}'\\ \mathbf{H}'\end{pmatrix}=\boldsymbol{\mathcal{D}}\cdot\begin{pmatrix}\mathbf{E}\\ \mathbf{H}\end{pmatrix},\qquad\qquad \boldsymbol{\mathcal{D}}=\frac{1}{\sqrt{2}}\begin{pmatrix}\mathbf{1} & \eta_0\mathbf{1}\\ -\eta_0^{-1}\mathbf{1} & \mathbf{1}\end{pmatrix}. \qquad (6)$$

The unprimed fields are associated with the nonreciprocal wedge structure, whereas the primed fields with the duality transformed wedge structure (see Fig. 7). The MHM-I and MHM-II are invariant under the considered duality transformation. In contrast, the matched gyrotropic material is mapped into a reciprocal material with an $\Omega$-coupling. The $\Omega$-medium is characterized by the 6×6 material matrix [18]

$$\mathbf{M}=\begin{pmatrix}\varepsilon_0\overline{\overline{\varepsilon}}_\Omega & -\dfrac{i\Omega}{c}\hat{\mathbf{z}}\times\mathbf{1}\\ -\dfrac{i\Omega}{c}\hat{\mathbf{z}}\times\mathbf{1} & \mu_0\overline{\overline{\mu}}_\Omega\end{pmatrix}. \qquad (7)$$

The anti-diagonal elements of **M** determine the magnetoelectric coupling tensors and the main diagonal elements the permittivity and permeability tensors of the material. Moreover, the constitutive parameters of the duality-transformed material satisfy $\Omega = i\varepsilon_{12} = \varepsilon_g$ and $\overline{\overline{\varepsilon}}_\Omega = \overline{\overline{\mu}}_\Omega = \frac{1}{2}\left(\overline{\overline{\varepsilon}} + \overline{\overline{\varepsilon}}^T\right)$, with $\overline{\overline{\varepsilon}}$ the permittivity of the gyrotropic material. Materials with an $\Omega$-coupling were originally introduced by Saadoun and Engheta [62-65] and recently received significant attention in the context of photonic topological insulators [12, 13] due to their potentials to guide electromagnetic energy with no backscattering [12, 13, 18]. It was shown in Ref. [18] that the TM-waves of the original problem (i.e., of the nonreciprocal wedge structure) are mapped into spin up ($\uparrow$) polarized modes with $E'_z = \eta_0 H'_z$. Similarly, the TE-polarized edge modes are transformed into spin down ($\downarrow$) polarized modes with $E'_z = -\eta_0 H'_z$.

Using the duality theory it is possible to transform the nonreciprocal wedge waveguide into a fully reciprocal structure that also behaves as energy sink. Indeed, by duality it is clear that the MHM-I interface of the $\Omega$-material wedge does not support edge waves. On the other hand, the MHM-II interface supports a spin $\uparrow$ edge mode that propagates towards the vertex with no back-reflections, and a spin $\downarrow$ edge mode that propagates away from the vertex. Therefore, albeit the structure is fully reciprocal, the vertex will behave as an energy sink of any the radiation generated near the MHM-II interface with spin $\uparrow$ polarization (see Fig. 7b). In contrast, the spin $\downarrow$ edge mode flows away from the sink, as shown in Fig. 7c. The spin $\uparrow$ and the spin $\downarrow$ waves are protected against backscattering from any perturbation that does not break the $\mathcal{P}\cdot\mathcal{T}\cdot\mathcal{D}$ symmetry [18]. In the present example, this property is a simple consequence of the duality link between the reciprocal system and the non-reciprocal gyrotropic system. The topological insulators of Khanikaev *et al* [12, 13] are particular cases of $\mathcal{P}\cdot\mathcal{T}\cdot\mathcal{D}$-

symmetric systems. Even though the considered structure is admittedly very challenging to realize, it can be in principle implemented using metamaterials [13, 66]. Furthermore, we believe that similar energy sinks can be observed in simpler reciprocal material platforms, but a detailed discussion is left for future work.

## IV. Conclusions

It was theoretically demonstrated that the formation of electromagnetic hotspots in nonreciprocal waveguides has a topological origin and is deeply rooted in the breakdown of the bulk edge correspondence in systems without a wave vector cut-off. Our analysis unveils that hotspots with a topological origin behave as "energy sinks" that absorb all the energy generated in their surroundings with no back-scattering. The electromagnetic field near the topological sinks can be ultra-singular. In the hypothetical limit of vanishing material loss the stored electromagnetic energy can grow steadily (linearly) with time when the sink is pumped by some external excitation. In realistic systems, the energy dragged towards the sink is dissipated in its neighborhood. Furthermore, using duality theory it was shown that similar energy sinks can be formed in reciprocal $\mathcal{P}\cdot\mathcal{T}\cdot\mathcal{D}$-invariant systems. We believe that these findings can be useful for energy harvesting and to enhance nonlinearities.

**Acknowledgements**

This work is supported in part by the IET under the A F Harvey Engineering Research Prize and by Fundação para Ciência e a Tecnologia (FCT) under project PTDC/EEITEL/4543/2014 and UID/EEA/50008/2019, and by the European Regional Development Fund (FEDER), through the Competitiveness and Internationalization Operational Programme (COMPETE 2020) of the Portugal 2020 framework, Project, RETIOT, POCI-01-0145-FEDER-016432. D. E. Fernandes acknowledges support by FCT, POCH, and the cofinancing of Fundo Social Europeu under the fellowship SFRH/BPD/116525/2016.

## Appendix A: The modal equation of the gyrotropic waveguide

Here we determine the exact the dispersion of the edge modes propagating in the waveguide depicted in Fig. 1. To this end, we write the electromagnetic field as a superposition of decaying exponentials (along the direction normal to the interface, i.e. along $\hat{\mathbf{y}}$). The magnetic field of the TM-waves is of the form:

$$H_z\left(x,y,\omega\right)=e^{ik_x x}\begin{cases} A_1 e^{-\gamma_G(y-d)} & y>d \\ A_2 e^{-\gamma_0 y}+A_3 e^{\gamma_0 y} & 0<y<d \\ A_4 e^{\gamma_m y} & y<0 \end{cases}, \qquad \text{(A1)}$$

with $k_x$ is the propagation constant of the edge mode along the *x*-direction, $\gamma_G=\sqrt{k_x^2-\varepsilon_{ef}\left(\omega/c\right)^2}$, $\gamma_m=\sqrt{k_x^2-\varepsilon_m\left(\omega/c\right)^2}$, $\gamma_0=\sqrt{k_x^2-\left(\omega/c\right)^2}$ are the attenuation constants along the *y*-direction in the gyrotropic material, metal and air, respectively. The unknowns $A_i$ (*i*=1,...,4) represent the amplitudes of the waves excited in the different regions. The electric field distribution is obtained from the Maxwell equations using $\mathbf{E}=\frac{1}{-i\omega}\frac{\overline{\overline{\varepsilon}}^{-1}}{\varepsilon_0}\cdot\nabla\times\mathbf{H}$. The tangential (with respect to the interface) electric field is given by:

$$E_x\left(x,y,\omega\right)=\frac{e^{ik_x x}}{-i\omega\varepsilon_0}\begin{cases} \frac{A_1}{\varepsilon_{11}^2+\varepsilon_{12}^2}\left(-\gamma_G\varepsilon_{11}+ik_x\varepsilon_{12}\right)e^{-\gamma_G(y-d)} & y>d \\ \gamma_0\left(-A_2 e^{-\gamma_0 y}+A_3 e^{\gamma_0 y}\right) & 0<y<d \\ A_4\frac{\gamma_m}{\varepsilon_m}e^{\gamma_m y} & y<0 \end{cases}. \qquad \text{(A2)}$$

and the normal component is equal to:

$$E_y\left(x,y,\omega\right)=\frac{e^{ik_x x}}{-i\omega\varepsilon_0}\begin{cases} \frac{A_1}{\varepsilon_{11}^2+\varepsilon_{12}^2}\left(-\gamma_G\varepsilon_{12}-ik_x\varepsilon_{11}\right)e^{-\gamma_G(y-d)} & y>d \\ -ik_x\left(A_2 e^{-\gamma_0 y}+A_3 e^{\gamma_0 y}\right) & 0<y<d \\ A_4\frac{-ik_x}{\varepsilon_m}e^{\gamma_m y} & y<0 \end{cases}. \qquad \text{(A3)}$$

The tangential components of the electromagnetic field are continuous at the interfaces $y=d$ and $y=0$. This yields a homogeneous system of equations of the form $\mathbf{B}\cdot\mathbf{v}=0$ with $\mathbf{v}=\begin{bmatrix}A_1 & A_2 & A_3 & A_4\end{bmatrix}^T$ and with the matrix $\mathbf{B}$ given by:

$$\mathbf{B}=\begin{bmatrix} 1 & -e^{-\gamma_0 d} & -e^{\gamma_0 d} & 0 \\ 0 & 1 & 1 & -1 \\ \dfrac{-\gamma_G\varepsilon_{11}+ik_x\varepsilon_{12}}{\varepsilon_{11}^2+\varepsilon_{12}^2} & \gamma_0 e^{-\gamma_0 d} & -\gamma_0 e^{\gamma_0 d} & 0 \\ 0 & -\gamma_0 & \gamma_0 & -\dfrac{\gamma_m}{\varepsilon_m} \end{bmatrix}. \quad \text{(A4)}$$

The edge mode dispersion is calculated by setting the determinant of $\mathbf{B}$ equal to zero, resulting in [21]:

$$\left(\frac{\gamma_G}{\varepsilon_{ef}}-\frac{\varepsilon_{12}ik_x}{\varepsilon_{11}^2+\varepsilon_{12}^2}\right)+\frac{\gamma_m}{\varepsilon_m}+\gamma_0\tanh\left(\gamma_0 d\right)+\left(\frac{\gamma_G}{\varepsilon_{ef}}-\frac{\varepsilon_{12}ik_x}{\varepsilon_{11}^2+\varepsilon_{12}^2}\right)\frac{\gamma_m}{\varepsilon_m\gamma_0}\tanh\left(\gamma_0 d\right)=0. \quad \text{(A5)}$$